\documentclass[preprint, 5p, authoryear]{elsarticle}
\usepackage{graphicx}

\begin{document}

\begin{frontmatter}

\title{How common are Earth-Moon planetary systems?}

\author[rvt]{S.˜Elser\corref{cor1}}
\ead{selser@physik.uzh.ch}

\author[rvt]{B.˜Moore}
\ead{moore@physik.uzh.ch}

\author[rvt]{J.˜Stadel}
\ead{stadel@physik.uzh.ch}

\author[focal]{R.˜Morishima}
\ead{ryuji.morishima@lasp.colorado.edu}

\cortext[cor1]{Corresponding author}

\address[rvt]{University of Zurich, Winterthurerstrasse 190, 8057 Zurich, Switzerland}
\address[focal]{LASP, University of Colorado, Boulder, Colorado 80303-7814, USA}

\begin{abstract}
The Earth's comparatively massive moon, formed via a giant impact on the proto-Earth, has played an important role in the development of life on our planet, both in the history and strength of the ocean tides and in stabilizing the chaotic spin of our planet. Here we show that massive moons orbiting terrestrial planets are not rare. A large set of  simulations by \citet{Morishima10}, where Earth-like planets in the habitable zone form, provides the raw simulation data for our study. We use limits on the collision parameters that may guarantee the formation of a circumplanetary disk after a protoplanet collision that could form a satellite and study the collision history and the long term evolution of the satellites qualitatively. In addition, we estimate and quantify the uncertainties in each step of our study. We find that giant impacts with the required energy and orbital parameters for producing a binary planetary system do occur with more than 1 in 12 terrestrial planets hosting a massive moon, with a low-end estimate of 1 in 45 and a high-end estimate of 1 in 4. 
\end{abstract}


\end{frontmatter}

\section{Introduction}

The evolution and survival of life on a terrestrial planet requires several conditions. A planet orbiting the central star in its habitable zone provides the temperature suitable for the existence of liquid water on the surface of the planet. In addition, a stable climate on timescales of more than a billion years may be essential to guarantee a suitable environment for life, particularly land-based life. Global climate is mostly influenced by the distribution of solar insolation \citep{Milan41, Berger84, Berger89, Atobe06}. The annual-averaged insolation on the surface at a given latitude is, beside the distance to the star, strongly related to the tilt of the rotation axis of the planet relative to the normal of its orbit around the star, the obliquity. If the obliquity is close to $0^{\circ}$, the poles become very cold due to negligible insolation and the direction of the heat flow is poleward. With increasing obliquity, the poles get more and more insolation during half of a year while the equatorial region becomes colder twice a year. If the obliquity is larger than  $57^{\circ}$, the poles get more annual insolation than the equator and the heat flow changes. Therefore, the equatorial region can even be covered by seasonal ice \citep{Ward00}. Thus, the obliquity has a strong influence on a planet's climate. The long-term evolution of the Earth's obliquity and the obliquity of the other terrestrial planets in the solar system, or planets in general, is controlled by spin-orbit resonances and the tidal dissipation due to the host star and satellites of the planet. Thus, the evolution of the planetary obliquity is unique for each planet.

Earth's obliquity fluctuates currently $ \pm\,1.3^{\circ}$ around  $23.3^{\circ}$ with a period of $ \sim41,000$ years \citep{Laskar93a,Laskar96}. The existence of a massive (or close) satellite results in a higher precession frequency which avoids a spin-orbit resonance. Without the Moon, the obliquity of the Earth would suffer very large chaotic variations. The other terrestrial planets in the solar system have no massive satellites. Venus has a retrograde spin direction, whereas a possibly initial more prograde spin may have been influenced strongly by spin-orbit resonances and tidal effects \citep{Goldreich70,Laskar96}. Mars' obliquity oscillates $ \pm 10^\circ$ degree around $ 25^\circ$ with a period of several $ 100,000$ years \citep{Ward74,Ward91}. Mercury on the other hand is so close to the sun that its rotation period is in an exact $ 3:2$ resonance with its orbital period. Mercury's spin axis is aligned with its orbit normal. 

On larger timescales, the variation of the obliquity can be even more dramatic. It has been shown that the tilt of Mars' rotation axis ranges from $ 0^\circ$ to $ 60^\circ$ in less than 50 million years and  $ 0^\circ$ to $ 85^\circ$ in the case of the obliquity of an Earth without the Moon \citep{Laskar93a,Laskar96}.

The main purpose of this report is to explore the giant impact history of the planets in order to calculate the probability of having a giant Moon-like satellite companion, based on simulations done by \citet{Morishima10}.  A giant impact between a planetary embryo called \textit{Theia}, the Greek titan that gave birth to the Moon goddess \textit{Selene}, first named by \citet{Halliday00},  and the proto-Earth is the accepted model for the origin of our Moon \citep{Hartmann75,Cameron76,Cameron91}, an event which took place within about $100\,Myr$ after the formation of calcium aluminum-rich inclusions in chondritic meteoroids, the oldest dated material in the solar system \citep{Touboul07}. After its formation, the Moon was much closer and the Earth was rotating more rapidly. The large initial tidal forces created high tidal waves several times per day, possibly promoting the cyclic replication of early bio-molecules \citep{Lathe04} and profoundly affecting the early evolution of life. Tidal energy dissipation has caused the Moon to slowly drift into its current position, but its exact orbital evolution is still part of an on-going debate \citep{Varga06,Lathe06}. Calculating the probability of life in the Universe \citep{Ward00} as well as the search for life around nearby planets may take into account the likelihood of having a massive companion satellite.

This report is structured as follows: In section \ref{sec:planetformation}, we give a brief review on the evolution of simulating terrestrial planet formation with N-body codes during the last decades and present the method we used. In section \ref{sec:satelliteformation}, we study the different parameters of a protoplanet collision to identify potential satellite forming events. In section \ref{sec:uncertainties}, we summarize the different uncertainties from the simulations and our analysis that may affect the final results. Finally, we give a conclusion, we present our results and compare them with previous works in section \ref{sec:results}.

\section{Simulating terrestrial planet formation}
\label{sec:planetformation}

There are good observational data on extra-solar gas giant planets, but whilst statistics on extra-solar rocky planets will be gathered in the coming years, for constraints on the formation of the terrestrial planets we rely on our own solar system. The established scenario for the formation of the Earth and other rocky planets is that most of their masses were built up through the gravitational collisions and interactions of smaller bodies \citep{Chamberlin05,Safranov69,Lissauer93}. \citet{Wetherill89} 
observed the phase of run-away growth. This phase is characterized by the rapid growth of the largest bodies. While their mass increases, their gravitational cross section increases due to gravitational focusing. When a body reaches a certain mass, the velocities of close planetesimals are enhanced, the gravitational focusing decreases and so does the accretion efficiency. This is called the oligarchic growth phase, first described by \citep{Kokubo98}. During this phase, the smaller embryos will grow faster than the larger ones. At the end, several bodies of comparable size are embedded in a planetesimal disk. These protoplanets merge via giant impacts to form the final planets. \citet{Dones93} showed that most of a terrestrial planet's prograde spin is imparted by the last major impactor and can not be accumulated via the ordered accretion of small planetesimals. Giant impacts with a certain impact angle and velocity generate a disk of ejected material around the target which is a preliminary step in the formation of a satellite. Usually, the simulations assume perfect accretion in a collision. Tables of the collision outcome can help to improve the simulations or to estimate the errors in the planetary spin, \citep{Kokubo10}.
Until recently simulations were limited in the number of planetesimal bodies that could be self-consistently followed for time spans of up to billions of years, but recent algorithmic improvements by \citet{Duncan98} in his SyMBA code and by \citet{Chambers99} in his Mercury code have allowed them to follow over long time spans a relatively large number of bodies ($\mathcal{O}(1000)$) with high precision, particularly during close encounters and mergers between the bodies, where individual orbits must be carefully integrated \citep{Chambers98,Agnor99,Raymond04,Kokubo06}. \citet{Raymond09} have also recently conducted a series of simulations where they varied the initial conditions for the gas giant planets and also track the accretion of volatile-rich bodies from the outer asteroid belt, leading either to ``dust bowl'' terrestrial planets or ``water worlds'' and everything between these extremes. All prior simulation methods with full interaction among all particles have however been limited in number of particles since their force calculations scale as $\mathcal{O}(N^2)$. 

We have developed a new parallel gravity code that can follow the collisional growth of planetesimals and the subsequent long-term evolution and stability of the resulting planetary system. The simulation code is based on an $\mathcal{O}(N)$ fast multipole method to calculate the mutual gravitational interactions, while at the same time following nearby particles with a highly accurate mixed variable symplectic integrator, which is similar to the SyMBA \citep{Duncan98} algorithm. Since this is completely integrated into the parallel code PKDGRAV2 \citep{Stadel01}, a large speed-up from parallel computation can also be achieved. We detect collisions self-consistently and also model all possible effects of gas in a laminar disk: aerodynamic gas drag, disk-planet interaction including Type-I migration, and the global disk potential which causes inward migration of secular resonances with gas dissipation. In contrast to previously mentioned studies, this code allows us to self-consistently integrate through the last two phases of planet formation with the same numerical method while using a large number of particles.

Using this new simulation code we have carried out 64 simulations which explore sensitivity to the initial conditions, including the timescale for the dissipation of the solar nebula, the initial mass and radial distribution of planetesimals and the orbits of Jupiter and Saturn \citep{Morishima10}. All simulations start with 2000 equal-mass particles placed between 0.5 and $4\,AU$. The initial mass of the planetesimal disk $m_{\rm d}$ is 5 or $10\,m_\oplus$. The surface density $\Sigma$ of this disk and of the initial gas disk depends on the radius through $\Sigma \propto r^{-p}$, where $p$ is 1 or 2. The gas disk dissipates exponentially in time and uniformly in space with a gas dissipation time scale $\tau_{\rm gas}=1, 2, 3$ or $5\,Myr$. After the disappearance of the gas disk (more precisely after time $\tau_{\rm gas}$ from the beginning of the simulation), Jupiter and Saturn are introduced on their orbits, e.g. circular orbits or the current orbits with higher eccentricities.

\begin{figure}

\includegraphics[scale=0.78]{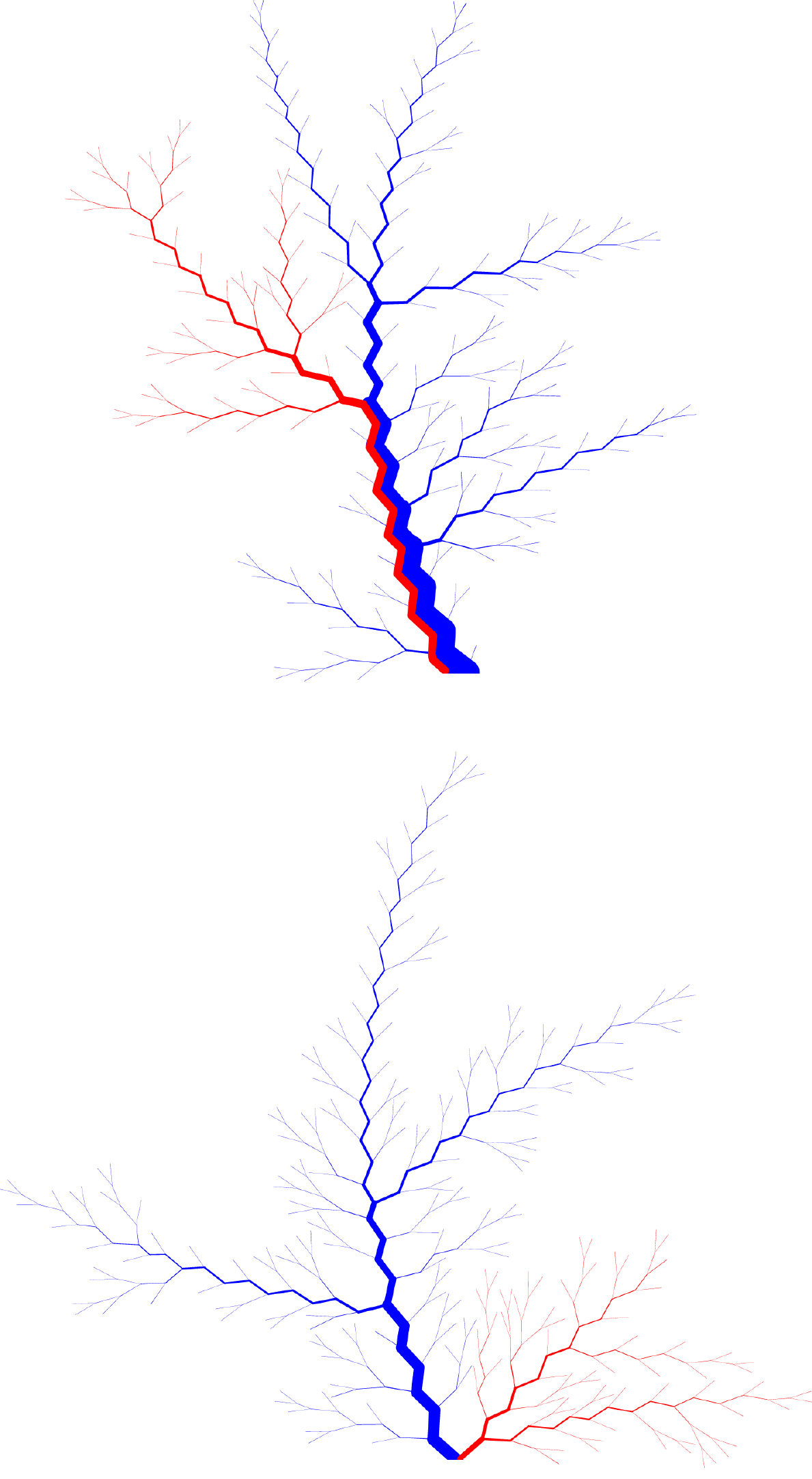}%
\centering
\caption{ Two merger trees.  They illustrate the accretion from the initial planetesimals to the last major impactors that merge  with the planet. Every 'knee' is a collision of two particles and the length between two collisions is given by the logarithm of the time between  impacts.  The thickness of the lines indicates the mass of the particle (linear scale). The red branch is the identified satellite forming impact in the planet's accretion history. \textit{Top:} a 1.1 $m_\oplus$ planet formed in the simulation with ($\tau_{\rm gas}$,p,$m_{\rm d}$)=(1\,Myr,2,10$\,m_\oplus$) and gas giants on circular orbits and a $0.3\,m_\oplus$ impactor. In this case, the moon forming impact is not the last collision event but it is followed by some major impacts. \textit{Right:}  a 0.7 $m_\oplus$ planet formed in the simulation with ($\tau_{\rm gas}$,p,$m_{\rm d}$)=(3\,Myr,1,5$\,m_\oplus$) and gas giants on the present orbits and a $0.2\,m_\oplus$ impactor. It is easy to see that the moon forming impact is the last major impact on the planet.  Although this planet is smaller than the upper one, it is composed out of a similar number of particles. In the case of the more massive disk ($m_{\rm d}=10$), the initial planetesimals are more massive since their number is constant. Therefore, fewer particles are needed to form a planet of comparable mass than in the case of $m_{\rm d}=5$.}
\label{fig:tree}%
\end{figure}

Figure \ref{fig:tree} shows two merger trees: a 1.8 $\,m_\oplus$ planet formed in the simulation with ($\tau_{\rm gas}$,p,$m_{\rm d}$)=(1\,Myr,1,5$\,m_\oplus$) and gas giants on the present orbits and a 1.1 $\,m_\oplus$ planet formed in the simulation with ($\tau_{\rm gas}$,p,$m_{\rm d}$)=(1\,Myr,2,10\,$m_\oplus$) and gas giants on circular orbits. The red branches are satellite forming impactors both with a mass $~0.3\,m_\oplus$ and represent two events of the final sample in figure \ref{fig:masses}. These merger trees with their different morphologies reveal the variety of collision sequences in terrestrial planet formation. They show that a large set of impact histories is generated by these simulations despite the relatively narrow parameter space for the initial conditions.

\section{Satellite formation}
\label{sec:satelliteformation}

During the last phase of terrestrial planet formation, the giant impact phase, satellites form. Collisions between planetary embryos deposit a large amount of energy into the colliding bodies and large parts of them heat up to several $10^3\,K$, e.g. \citet{Canup04}. Depending on the impact angle and velocity and the involved masses, hot molten material from the target and impactor can be ejected into an circumplanetary orbit. This forms a disk of ejecta, the disk material is in a partially vapor or partially molten state, around the target planet. The proto-satellite disk cools and solidifies. Solid debris form and subsequently agglomerate into a satellite \citep{Ohtsuki93,Canup96,Kokubo00}. 

The giant impact which resulted in the Earth-Moon system is a very particular event \citep{Cameron91,Canup04}. The collision parameter space that describes a giant impact can by parametrized by $\gamma=m_i/m_{\rm tot}$, the ratio of impactor mass $m_i$ to total mass in the collision $m_{\rm tot}$, by $v\equiv v_{\rm imp}/v_{\rm esc}$, the impact velocity in units of the escape velocity $v_{\rm esc}=\sqrt{2Gm_{\rm tot}/(r_i+r_t)}$, where $r_t$ and $r_i$ are the radii of target and impactor. Furthermore, it is described by the scaled impact parameter $b$, where $b=0$ indicates a head-on collision and $b=1$ a grazing encounter, and the total angular momentum $L$. Recent numerical results \citep{Canup08} obtained with smoothed particle hydrodynamic (SPH) simulations for the Moon-forming impact parameters require: $\gamma\sim0.11$, $v\sim1.1$, $b\sim0.7$ and $L\sim1.1\,L_{\rm EM}$, where $L_{\rm EM}$ is the angular momentum of the present Earth-Moon system. These simulations also include the effect of the initial spins of the colliding bodies, but the explored parameter space is restricted to being close to the Moon-forming values given above. 

If one does not focus on a strongly constrained system like the Earth-Moon system but just on terrestrial planets of arbitrary mass with satellites that tend to stabilize their spin axis, the parameter space is broadened. It becomes difficult to draw strict limits on the parameters because collision simulations for a wider range of impacts were not available for our study. Hence, based on published Moon-forming SPH simulations by Canup (2004, 2008), we use a semi-analytic expression to constrain the mass of a circumplanetary disk that can form a satellite. In addition, we include tidal evolution and study the ability of the satellite to stabilize the spin axis of the planet.

\subsection{Satellite mass and collision parameters}

We do not know the exact outcome of a protoplanet collision, but certainly the satellite mass is related to the collision parameters of the giant impact. Hence, we can draw a connection from these parameters to the mass of the final satellite. Based on the studies of the Earth's Moon formation, we can start with a simple scaling relation: a Mars-size impactor gives birth to a Moon-size satellite. Their mass ratio is $ m_{\rm Mars}/m_{\rm Moon}\sim 10$. Thus, to very first approximation, we can assume that an impactor mass is usually 10 times larger than the final mass of the satellite. Of course, this shows that only a small amount of material ends up in a satellite, but this statement is only valid for a certain combination of mass ratio, impact parameter and impact speed and usually gives an upper limit on the satellite mass.

In order to get a better estimation of the satellite mass, we use the method obtained in the appendix of \citet{Canup08}. 

There, an expression is derived that describes the mass of the material that enters the orbit around the target after a giant impact:
\begin{equation}
\frac{m_{\rm disk}}{m_{\rm tot}}\sim C_{\gamma}\left(\frac{m_{\rm pass}}{m_{\rm tot}}\right)^2
\label{eq:diskmass}
\end{equation}
where the prefactor $C_{\gamma}\sim2.8\left({0.1}/{\gamma}\right)^{1.25}$ has been determined empirically from the SPH data. $m_{\rm pass}/m_{\rm tot}$ is mass of the impactor that avoids direct collision with the target. It depends mainly on the impact parameter $b$ and on the mass ratio $\gamma$ and can be computed by studying the geometry of the collision. The total impactor volume that collides with the target is:
\begin{equation}
V_T=\int_0^{\pi} A(\phi) \, d\phi,
\label{eq:volume}
\end{equation}
with
\begin{equation}
A(\phi)=r_t^2\theta_t(\phi)+r_i^2\theta_i(\phi)-D r_i\sin\phi \sin\theta_i(\phi),
\end{equation}
where $r_i$ and $r_t$ are impactor and target radius, $D=b(r_i+r_t)$ gives the distance between the centers of the bodies and

\begin{equation}
\theta_i(\phi)=cos^{-1}\left[\frac{D^2+r_i^2\sin^2\phi-r_t^2}{2Dr_i\sin\phi}\right],
\end{equation}
\begin{equation}
\theta_t(\phi)=cos^{-1}\left[\frac{D^2+r_t^2-r_i^2\sin^2\phi}{2Dr_t}\right].
\end{equation}

Assuming a differentiated impactor with $r_{\rm core}\sim 0.5 r_i$ and repeating the above integration for this radius, the colliding volume of the impactor mantle is $V_{\rm mantle}=V_T - V_{\rm core}$. If we assume that the core is iron and the mantle dunite, the core density $\rho_{\rm core}$ has  roughly twice the density of the mantle $\rho_{\rm mantle}$. The mass of the impactor that hits the target is $m_{\rm hit}=\rho_{\rm core}V_{\rm core}+\rho_{\rm mantle}V_{\rm mantle}$. Therefore, the mass that passes the target is $m_{\rm pass}=m_i-m_{\rm hit}$.
 
\begin{figure}
\centering%
\includegraphics[scale=0.6]{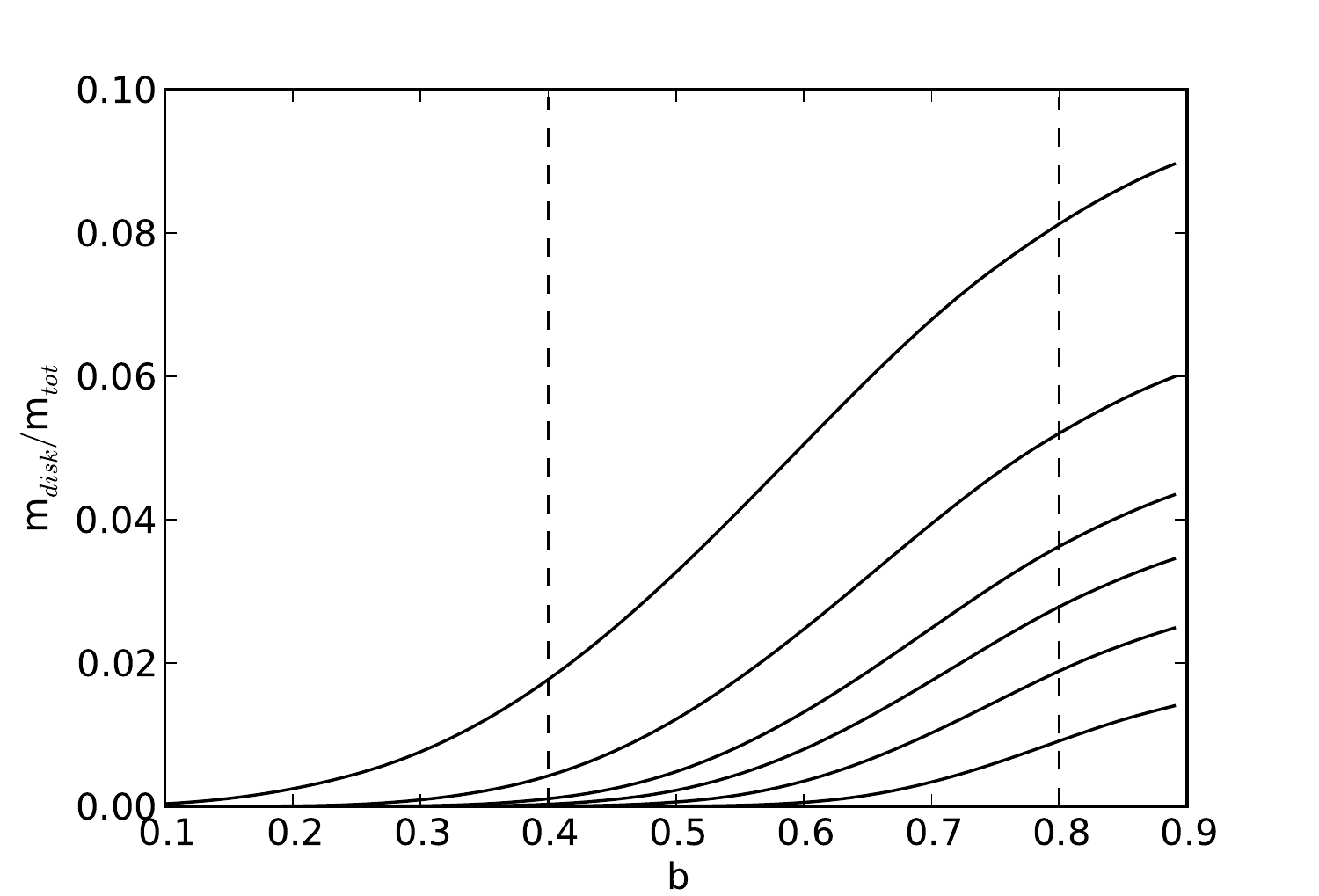}%
\caption{ The disk mass resulting after a giant impact in units of the total mass of the colliding system relative to the impact parameter $ b$ based on equation (\ref{eq:diskmass}). The different solid lines belong to different mass ratios $ \gamma$. From bottom to top: $\gamma= 0.05, 0.1, 0.15, 0.2, 0.3$ and 0.5. These mass ratios result in different disk masses. The equation is valid up to a factor 2 in between the two dash lines ($0.4<b<0.8$) for small velocities. The disk mass is an upper bound on the satellite mass.}
\label{fig:diskb}%
\end{figure}

We use the full expression derived by \citet{Canup08}, equation (\ref{eq:diskmass}), to estimate the disk mass resulting from the giant impacts in our simulation. Equation (\ref{eq:diskmass}) is correct to within a factor 2, if $ v<1.4$ and $ 0.4<b<0.7$ or if $ v<1.1$ and $ 0.4<b<0.8$. Figure \ref{fig:diskb} illustrates the amount of material that is transported into orbit for the parameter range $ 0.4<b<0.8$ (see figure B2 in \citet{Canup08} for more details). It shows that a small impact parameter $b$ reduces the material ejected into orbit significantly. The same holds for a reduction of $\gamma$, because those collisions are more grazing. Based of simple arguments, we use the limits above to identify the moon forming collision in the $(v,b)$-plane. Details on how this assumption affects the result are given in section \ref{sec:uncertainties}.

\citet{Ida97} and \citet{Kokubo00} studied the formation of a moon in a circumplanetary disk through N-body simulations. They found that the final satellite mass scales linearly with the specific angular momentum of the disk. The fraction of the disk material that is finally incorporated into the satellite ranges form 10 to $55 \%$. Thus, we assume that not more than half of the disk material is accumulated into a single satellite. The angular momentum of the disk is unknown and we can not use the more exact relationships.

\subsection{Spin-orbit resonance}

Spin-orbit resonance occurs when the spin precession frequency of a planet is close to one of the planet's orbital precession frequencies. It causes large variation in the obliquity \citep{Laskar96}, the angle between this spin axis and the normal of the planet's orbital plane. An obliquity stabilizing satellite increases the spin precession frequency to a non-resonant (spin-orbit) regime. To ensure this, one can set a rough limit on the system parameters \citep{Atobe04} through 
\begin{equation}
\frac{m_s}{a_s^3}\gg\frac{m_\ast}{a_p^3},
\label{eq:atobe04}
\end{equation}
 where $m_s$ is the mass of the satellite, $m_{\rm \ast}$ the mass of the central star and $a_s$ the semi-major axis of the satellite's orbit and $a_p$ the semi-major axis of the planet. 
 
If the left term of the inequality is much larger than the right one, the spin precession frequency of the planet should be high enough to ensure that it is over the upper limit of the orbital precession frequency so that spin-orbit resonance does not occur. 
Although this inequality is very simplified, we try to estimate the minimum mass of a satellite such that it is able to stabilize the obliquity of its planet.

The exact semi-major axis of a satellite after formation is unknown but the Roche limit is the lower bound of its semi major axis.  The Roche limit  $a_{\rm R}$ of the planet is \citep{Murray99}:
\begin{equation}
a_{\rm R} = r_{ s} \left(\frac{3m_{ p}}{m_{ s}}\right)^{\frac{1}{3}},
\label{eq:roche}
\end{equation}
where $ r_s$ is the radius of the satellite and $ m_s$ its mass and the mass of the planet is given by $ m_p$. \citet{Ohtsuki93} and \citet{Canup96} provided detailed analytic treatments of the accretion process of satellites in an impact-generated disk. Based on those studies, \citet{Kokubo00} have shown that the true value of the radius of satellite accretion will not diverge much from the Roche radius in the case of the Earth-Moon system, a typical satellite orbit semi-major axis in their simulations was $a \simeq 1.3\,a_{\rm R}$. We used this approximation to estimate a lower bound on the satellite-planet mass ratio. We rewrite the Roche limit as 
\begin{equation}
a_{\rm R}=\left(\frac{3}{2}\right)^{\frac{2}{3}}\left(\frac{m_{ p}}{\pi \rho_{ p}}\right)^\frac{1}{3}
\label{eq:roche1}
\end{equation}
where we used $ r_{ s}^3=(\frac{4\pi}{3}\rho_{ s})^{-1}m_s$ and the fact that $ \rho_{ p}=\rho_{ s}$ in our model. We insert this in equation (\ref{eq:atobe04}) instead of $ a_s$ and get the condition
\begin{equation}
\frac{ m_s}{m_p}\gg\frac{9m_{\ast}}{4\pi \rho_p a_p^3}.
\label{eq:ms/mp}
\end{equation} 
Inserting a density of  $2\,g\,cm^{-2}$, the density of the bodies in the Morishima simulations, and planet semi-major axis of $1\,AU$, we get a mass ratio of $ \sim 10^{-5}$, which is smaller than the minimum ratio of the smallest and largest particles in our simulations. This is a lower limit on the stabilizing satellite mass but the tidal evolution of the planet-satellite system can alter this limit dramatically.

The orbital precession frequencies of a planet depend on the neighbouring or massive planets in its system. In the case of the Earth, Venus, Jupiter and Saturn cause the most important effects. To keep the spin precession frequency high enough, a spin period below 12 h would have the same effect as the present day Moon \citep{Laskar93a}. Hence, even without a massive satellite, obliquity stabilization is possible as long as the planet is spinning fast enough. However, the moon forming impact provides often a significant amount of angular momentum. A more sophisticated analysis including the precession frequencies of the all planets involved or formed in the simulations and a better treatment of the collisions to provide better estimates of the planetary spins would clearly be an improvement but this is out of the scope of this work.

\subsection{Tidal evolution}

After its formation, even a small satellite is stabilizing the planet's obliquity. Its fate is mainly controlled by the spin of the planet, the orientation of the spin axis of the planet relative to its orbit around the central star and relative to the orbital plane of the satellite and by possible spin-orbit resonance. Which satellites will continue to stabilize the obliquity as they recede from the planet? Orbital evolution is a complicated issue \citep{Atobe06} and there is still an ongoing debate even in the case of the Earth-Moon system, as mentioned. To classify the different orbital evolutions, it is helpful to introduce the 
synchronous radius, at which a circular orbital period equals the rotation period of the planet. In the prograde case, a satellite outwards of the synchronous radius will recede from the planet as angular momentum is tidally transferred from the planet to the satellite and the spin frequency of the planet decreases. In this case the synchronous radius will grow till it equals the satellite orbit. Angular momentum is transferred faster if the mass of the satellite is large, since the tidal response in the planet due to the satellite is greater. Large mass satellites will quickly reach this final co-rotation radius, where their recession stops. Even though their orbital radius becomes larger, these moons are massive enough to satisfy inequality (\ref{eq:atobe04}) and avoid spin-orbit resonances. Small satellites will recede very slowly compared to their heavy brothers and eventually fulfill the condition (\ref{eq:atobe04}) within the host star's main sequence life time. In contrast, low mass satellites that form in situ far outside the Roche limit, should this be possible, will probably not stabilize the spin axis. Intermediate mass satellites with $ m_s\sim m_{\rm Moon}$ may recede fast enough so that they tend to lose their obliquity stabilizing effect during a main sequence life time. The Earth-Moon system shows that even in this intermediate mass regime, long term stability can occur, since the Moon has stabilized the Earth's obliquity for billion of years. 

\begin{figure}
\centering%
\includegraphics[scale=0.8]{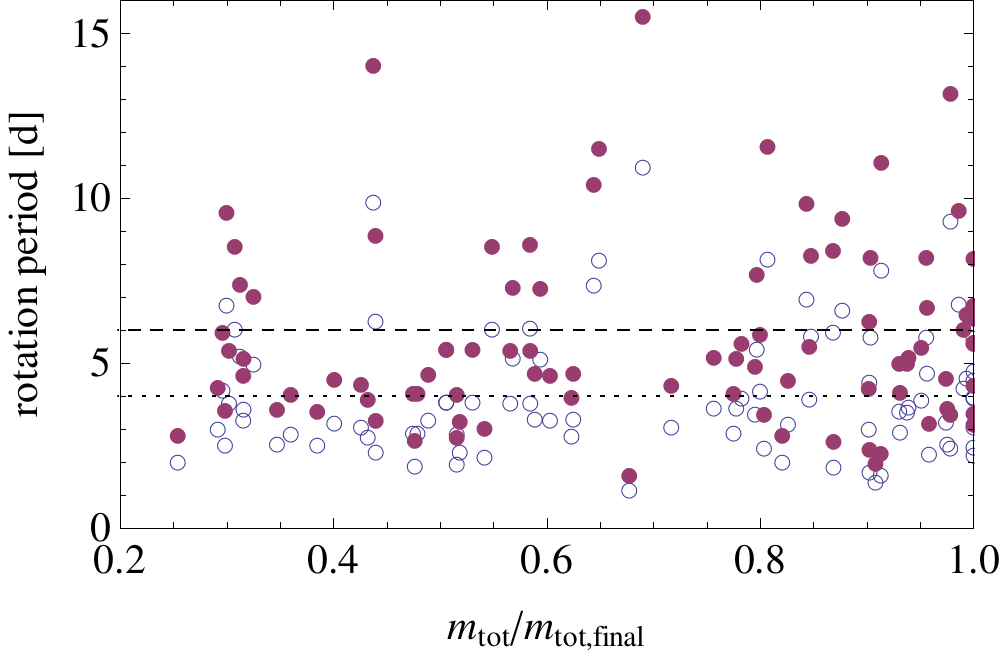}%
\caption{The mass of the planet in units of its final mass $m_{\rm tot}/m_{\rm tot,final}$ relative to the rotational period in days of the body after the moon-forming collision. This is the final sample but without excluding moon formation inside the synchronous radius. \textit{Empty circles:} The decrease of the planet spin due to escaping material or the  formation of a satellite are not included in the calculation of the rotation period, it is based on the perfect accretion assumption. \textit{Full circles:} The angular momentum is assumed to be $30\%$ smaller due to a more realistic collision model \citep{Kokubo10}. The dotted line gives the position of the threshold for synchronous rotation for the Roche radius $a_{\rm R}$, the dashed line gives its position for the $1.3\,a_{\rm R}$. We want to focus on the better estimate of the anuglar momentum (full circles) and the initial moon radius (dashed line). Hence, roughly a quarter of all moon forming collisions are excluded in the final sample, figure \ref{fig:masses}. }
\label{fig:rotation}%
\end{figure}

On the other hand, a satellite inside the synchronous orbit will start to spiral towards the planet, while its angular momentum is transferred to the planet. Soon, it is disrupted by tidal forces or will crash on the planet. How do we know if a satellite forms inside or outside the synchronous radius? The Roche limit (\ref{eq:roche}) depends on the mass and size of the bodies while the synchronous radius $r_{\rm sync}$ is a function of the planet mass $m_p$ and depends inversely proportional on its rotation frequency, which is obtained from equating the gravitational acceleration and the centripetal acceleration:
\begin{equation}
r_{\rm sync}=(Gm_p)^\frac{1}{3}\omega_p^{-\frac{2}{3}}.
\end{equation}
Equating this formula with (\ref{eq:roche}) gives:
\begin{equation}
\left(\frac{3}{2}\right)^{\frac{2}{3}}\left(\frac{m_{ p}}{\pi \rho_{ p}}\right)^\frac{1}{3}=(Gm_p)^\frac{1}{3}\omega_p^{-\frac{2}{3}}.
\end{equation}
The planet mass drops out and we get an lower limit on the planet angular velocity to guarantee a satellite outside the Roche radius:
\begin{equation}
\omega_{p,min}=\frac{2}{3}(\pi G \rho_p)^{\frac{1}{2}}=0.00043\,s^{-1},
\end{equation}
which equals a rotation period of $4\,$h.
The final planets of \citet{Morishima10} have generally a high rotational speed, some of them rotating above break up speed. The rotation period after the moon forming collision is usually around 2-5 hours, see figure \ref{fig:rotation}.  A more exact particle growth model without the assumption of perfect sticking would lower the rotation frequency by roughly $30\%$ \citep{Kokubo10}, where we can also include the loss of rotational angular momentum due to satellite formation (full circles). On the other hand, the mean distance of satellite formation is around $1.3\,a_{\rm R}$, and the maximum rotation period of the planet for a receding satellite changes to $\sim6\,h$ (dashed line). Hence, applying this to our final sample, 1/4 of all moon forming collisions are excluded.

Finally, for the remaining events, we assume that the planet spin is large enough so that the synchronous radius is initially smaller than the Roche radius. Satellites that form behind the Roche limit will start to recede from the planet. We can conclude that almost every satellite-planet system in our simulation will fulfill (\ref{eq:atobe04}). % To include satellite orbit evolution more properly connected with spin orbit resonance in each simulation could be a future step.

A special outcome of the tidal evolution of a prograde planet-satellite system is described by \citet{Atobe06}. If the initial obliquity $\theta$ of the planet after the moon forming impact is large, meaning that the angle between planet spin axis and planet orbit normal is close to $ 90^\circ$, results in a very rapid evolution when compared to the previously discussed case of a moon receding to the co-rotating radius and becoming tidally locked. The spin vectors of the protoplanets are isotropically distributed after the giant impact phase \citep{Agnor99} and the obliquity distribution corresponds to $ p(\theta)=\frac{1}{2} \sin(\theta)$. In the most extreme scenario, a massive prograde satellite will crash onto the planet in a timescale of order 10,000 years after formation, even though it initially recedes. Hence, without a favorable initial obliquity is an added 
requirement for the survivability of a close or massive moon. We include this by discarding massive and highly oblique impacts by an approximated inequality, based on area B in figure 13 in \citet{Atobe06}:
\begin{equation}
\theta<\frac{\pi}{2}-\frac{\pi}{0.2}\frac{m_s}{m_p} 
\end{equation}
More exactly, this holds best for planets with $ a_p\sim1\,AU$, and it has only to be taken into account if $ {m_s}/{m_p}<0.05$. The distribution of the $a_p$ of the simulated planets ranges for 0.1 to $4\,$AU. In the case of a smaller distance to the host star, angular momentum is removed faster from the planet-satellite system and the evolution timescales are shorter in general. To use this limit properly, a more general expression for different semi-major axes has to be derived, which is out of the scope of this work. Moreover, it depends linearly on the satellite mass which is overestimated in general. A smaller satellite mass would reduce the number of excluded events in general. However, when applying this constraint on the data, only one out of seven moon forming collisions are affected. Since it is over-simplified, we exclude it from our analysis.   

For retrograde impacts, where the impactor hits in opposition to the target's spin, two cases result in differing evolution. 
If the angular momentum of the collision is much larger than the initial rotational angular momentum, the spin direction of the planet is reversed and any impact-generated disk will rotate in the same direction as the planet's spin. 
On the other hand, a retrograde collision with small angular momentum will not alter the spin direction of the planet significantly and it becomes possible to be left with a retrograde circumplanetary disk\citep{Canup08}. After accretion of a satellite, tidal deceleration due to the retrograde protoplanet spin will reduce the orbital radius of the bodies continuously till they merge with the planet. Hence, a long-lived satellite can hardly form in this case.

In order to find a reasonable threshold between these two regimes based on the limited information we have, we assume that the sum of the initial angular momenta of the bodies $\vec L_{\rm t}$ and $\vec L_{\rm i}$ and of the collision $\vec L_{\rm col}$ equals the spin angular momentum of the planet and the satellite  $\vec L_{\rm planet}$ and $\vec L_{\rm moon}$ after the collision plus the angular momentum of the orbiting satellite $\vec L_{\rm orbit}$ at $1.3\,a_{\rm R}$ parallel to the collision angular momentum. 

 Comparing initial and final angular momenta gives:
\begin{equation}
\vec L_{\rm col} + \vec L_{\rm t} +\vec L_{\rm i}=\vec L_{\rm orbit}+\vec L_{\rm moon}+\vec L_{\rm planet},
\label{eq:Lequ}
\end{equation}
where $\vec L_{\rm orbit}$ is parallel to $\vec L_{\rm col}$: 
\begin{equation}
\vec L_{\rm orbit}=|\vec L_{\rm orbit}|\frac{\vec L_{\rm col}}{|\vec L_{\rm col}|}. 
\end{equation}

We assume that $|\vec L_{\rm moon}|/|(\vec L_{\rm orbit}+\vec L_{\rm planet})| \ll 1$ and
\begin{equation}
L_{\rm orbit}=m_s a_s^2 n=m_s a_s^2 \sqrt{\frac{G m_p}{a_s^3}}=m_s\sqrt{G a_s m_p},
\end{equation}
where $n$ is the orbital mean motion of the satellite if its mass is much smaller than the planet mass, $n\sim\sqrt{G m_{p}/a_s^3}$, and G is the gravitational constant. With $a_s=a_{R}$ we get
\begin{equation}
L_{\rm orbit}=m_s (G m_p)^\frac{1}{2}\left(\frac{3}{2\pi}\frac{1}{\rho_p}m_p\right)^\frac{1}{6}.
\end{equation}
If 
\begin{equation}
\frac{|\vec L_{\rm col} + \vec L_{\rm t}|}{\sqrt{(L_{\rm col}^2 + L_{\rm t}^2)}}<1,
\end{equation}
the collision is retrograde. Hence,
\begin{equation}
\vec L_{\rm planet}=\vec L_{\rm col} + \vec L_{\rm t} +L_{\rm i}-|\vec L_{\rm orbit}|\frac{\vec L_{\rm col}}{|\vec L_{\rm col}|},
\end{equation} 
If $\vec L_{\rm planet}$ and $\vec L_{\rm orbit}$, which is parallel to $\vec L_{\rm col}$, are retrograde,
\begin{equation}
\frac{|\vec L_{\rm col} + \vec L_{\rm orbit}|}{\sqrt{(L_{\rm col}^2 + L_{\rm orbit}^2)}}<1,
\end{equation}
the satellite will be tidally decelerated. Those cases are excluded from being moon-forming events. Spin angular momenta are in general overestimated since material is lost during collisions in general, but as mentioned above, the simulations of \citet{Morishima10} assume perfect accretion. We include this consideration by reducing the involved spins and the orbit angular momentum of the satellite by $30\%$, \citep{Kokubo10}. 

Both scenarios described above, a large initial obliquity or a retrograde orbiting planet, might not become important until subsequent impactors hit the target. Giant impacts can change the spin state of the planet in such a way that the satellite's fate is to crash on the planet. This scenario is discussed in the next section.

\subsection{Collisional history}
We exclude all collisions from being satellite-forming impacts whose target is not one of the final planets of a simulation. A satellite orbiting an impact is lost through the collision with the larger target. 

\begin{figure}
\centering%
\includegraphics[scale=0.8]{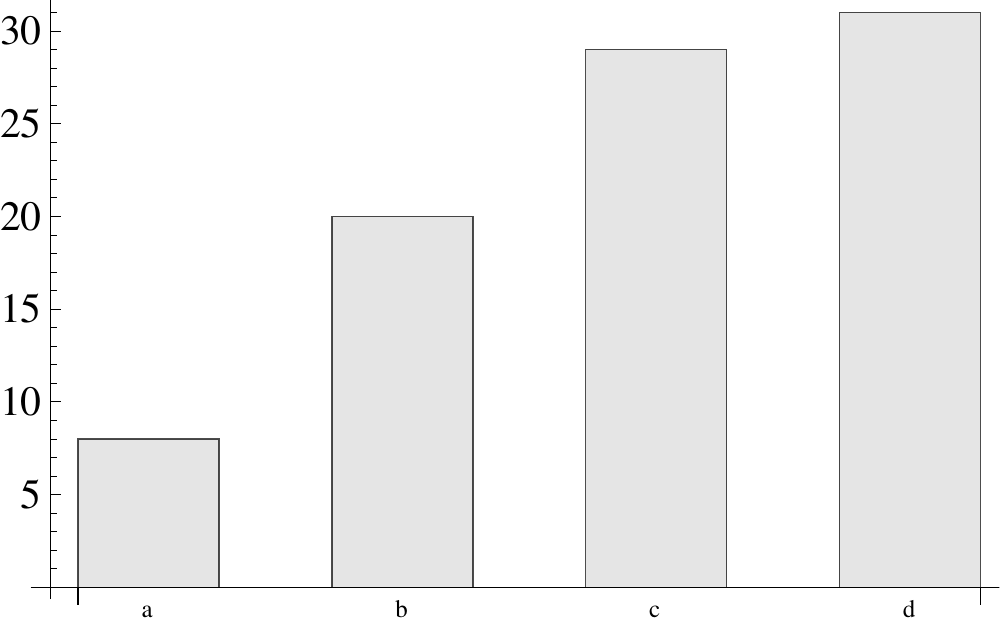}%
\caption{This bar chart shows the distribution of the moon forming collisions in four groups with different impact history with respect to the last major impact. a: the moon forming impact is the last major impact on the planet. b: there are multiple moon forming impacts, but the last one is also the last major impact. c: there is only one moon forming impact and it is followed by subsequent major impacts. d: there are multiple moon forming impacts, but the last of them is followed by subsequent major impacts.}
\label{fig:bars}%
\end{figure}

Multiple giant impacts occur during the formation process of a planet and it is useful to study the impact history in more detail. Subsequent collisions and accretion events on the planet after the satellite-forming event may have a large effect on the final outcome of the system. We choose a limit of $5 m_{\rm planetesimal}$ to distinguish between large impacts and impacts of small particles, which are responsible for the ordered accretion. To stay consistent, the same limit is used below to exclude small impactors from our analysis. We divide all identified moon forming events into four groups:
\begin{itemize}
	\item[a)] The moon forming event is the last major impact on the planet. Subsequent mass growth happens basically through planetesimal accretion (see merger tree at the bottom in figure \ref{fig:tree}).
	\item[b)] There are several moon forming impacts, in which the last impact is the last major impact on the planet.
	\item[c)] The moon forming event is not the last giant impact on the planet. The satellite can be lost due to a disruptive near or head-on-collision \citep{Stewart99} of an impactor and the satellite. A late giant impact on the planet can change the spin axis of the planet and the existing satellite can get lost due to tidal effects (see tree at the top in figure \ref{fig:tree}).
	\item[d)] There are several moon forming events in the impact history of the planet, followed by additional major impacts. As before, the moon forming collisions can remove previously formed satellites. On the other hand, and existing moon can have an influence on the circumplanetary disk formed by a giant impact and can suppress the formation of multiple satellites orbiting the planet.
\end{itemize}
The final states of the planet-satellite systems in group $c$ are difficult to estimate, since such systems might change significantly by additional giant impacts. To a lesser extent, this holds for $d$, but those systems are probably more resilient to the loss of satellites by direct collisions. The number of events per group is shown in figure \ref{fig:bars}. Group $c$ and $d$ include more then 2/3 of all collisions. Group $c$ is the most uncertain and we use it to quantify the error on the final sample. 

\begin{figure}
\centering%
\includegraphics[scale=0.8]{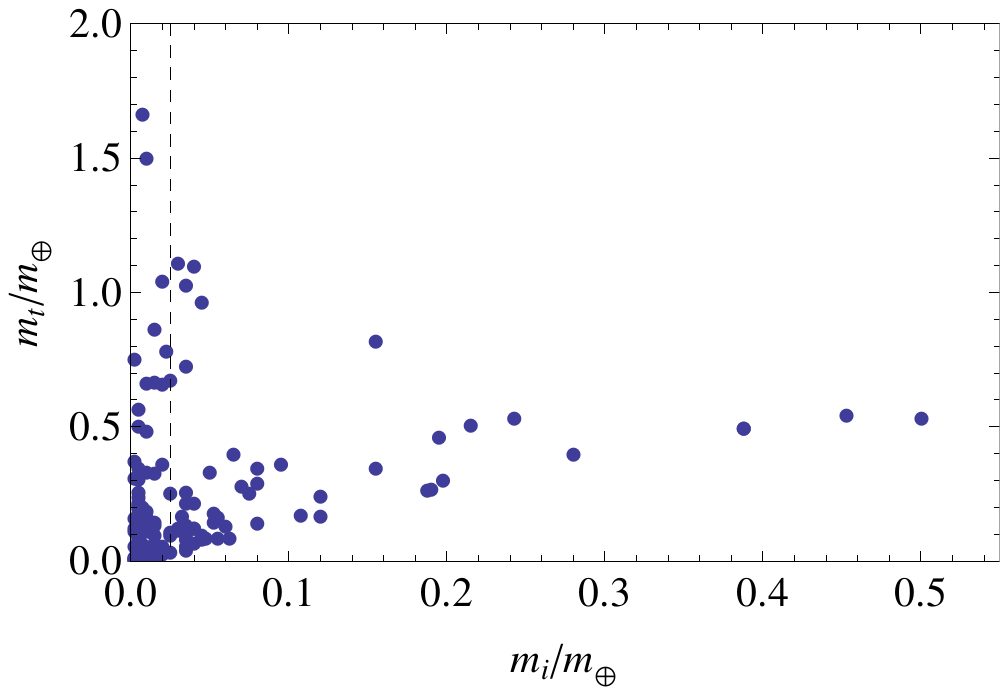}%
\caption{ The target mass $ m_t$ versus the impactor mass $ m_i$ involved in the satellite-forming collisions. We exclude events that include impactors and targets with masses of the order of the initial planetesimal mass ($  m_i, m_t < 5\, m_{\rm planetesimal}$) indicated by the dashed line to avoid resolution effects. The line in the case of the target mass is not shown since it is very close the frame. Due to this cut, the total number of satellites decreases significantly while the number of massive satellites in our analysis increases. The shown sample is the final set of events without applying the threshold for small particles. Therefore, the number of accepted events in this plot does not equal the number of satellites in figure \ref{fig:masses}, since this cut is not applied. If the cut is used, new events are accepted, which where neglected before because they were not the last moon forming impacts on the target.}
\label{fig:targimp}%
\end{figure}

%A more practicle exclude all collisions with targets smaller than half of a final planet mass from being moon forming events. In figure \ref{fig:fractime}, we see that a significant amount of satellite-forming events take place in the early stages of planet formation, resulting in satellites which probably do not survive the subsequent formation process. This limit is artificial but a simple way to include this probability. Without this limit, the number of satellites will increase by a factor of 1.5.
Furthermore, we exclude impactors and targets that have masses of the order of an initial planetesimal mass ($5 \,m_{\rm planetesimal} \sim 0.0025\,m_\oplus$) from producing satellite forming events since their masses are discretized and related to the resolution of the simulation. In addition, small impactors will probably not have enough energy to eject a significant amount of material into a stable orbit. Therefore, setting a lower limit on the target and impactor mass results in the exclusion of many collisions but not in a significant underestimation of the true number of satellites, see figure \ref{fig:targimp}.

\section{Uncertainties}
\label{sec:uncertainties}
Our final result depends on several assumptions, limitations and approximations. In this section we want to quantify them as much as possible and summarize them.

The data of \citet{Morishima10} we are using has two peculiarities worth mentioning: The focus on the Solar System and the small number of simulations per set of initial conditions. 

The simulations were made in order to reproduce the terrestrial planets of the Solar System. The central star has 1 solar mass and the two gas giants that are introduced after the gas dissipation time scale have the mass of Jupiter and Saturn and the same or similar orbital elements. Although the initial conditions like initial disk mass or gas dissipation time scale are varied in a certain range, the simulated systems do not represent general systems with terrestrial planets. However, we assume that the 
range of impact histories is representative of the range that would be seen in other systems. Merger trees (figure \ref{fig:tree}) reveal that those simulations cover a huge diversity of impact histories. But future work will need to investigate the full 
range of impact histories that could be relevant to the formation of terrestrial planets in other, possibly more exotic, extra-solar systems.

However, the set of simulation covers a broad range of initial conditions. But for every set of initial conditions, only one simulation exists since they are very time consuming (each simulation requires about 4 months of a quad-core CPU). Therefore, it is difficult to separate effects of the choice of certain initial parameters from effects of stochastic processes. We grouped our moon forming events with respect to the simulation parameter in question. The only parameter that reveals an effect on the final sample is the gas dissipation time scale $\tau_{\rm gas}$. Larger time scales lead to less moon forming collisions. This variation is correlated with mass and number of final planets. If the gas disk stays for several million years, the bodies are affected by the gas drag for a long time, spiral towards the sun and get destroyed. Therefore, there are less giant impacts and smaller planets. One would suppose that the initial mass of the gas disk should be correlated to the mass of the final planets and therefore to the number of giant impacts, but the two initial protoplanetary disk masses of $5\,m_{\oplus}$ and $10\,m_{\oplus}$ show no essential difference.

The approach we use to identify events and estimate the mass of the satellite is also based on various approximations and limitations.

\paragraph{Satellite mass}
The method we use to calculate the mass of the circumplanetary disk is valid to better than a factor of 2 within the parameter range we use \citep{Canup08}. Only 10-55$\%$ of the disk mass are embodied in the satellite \citep{Kokubo00}. Therefore, the mass we use is just half of the estimated disk mass, and in the worst case, the satellite is ten times less massive than estimated. This uncertainty affects the number of massive satellites but not the number of satellites in general.

\begin{figure}
\centering%
\includegraphics[scale=0.9]{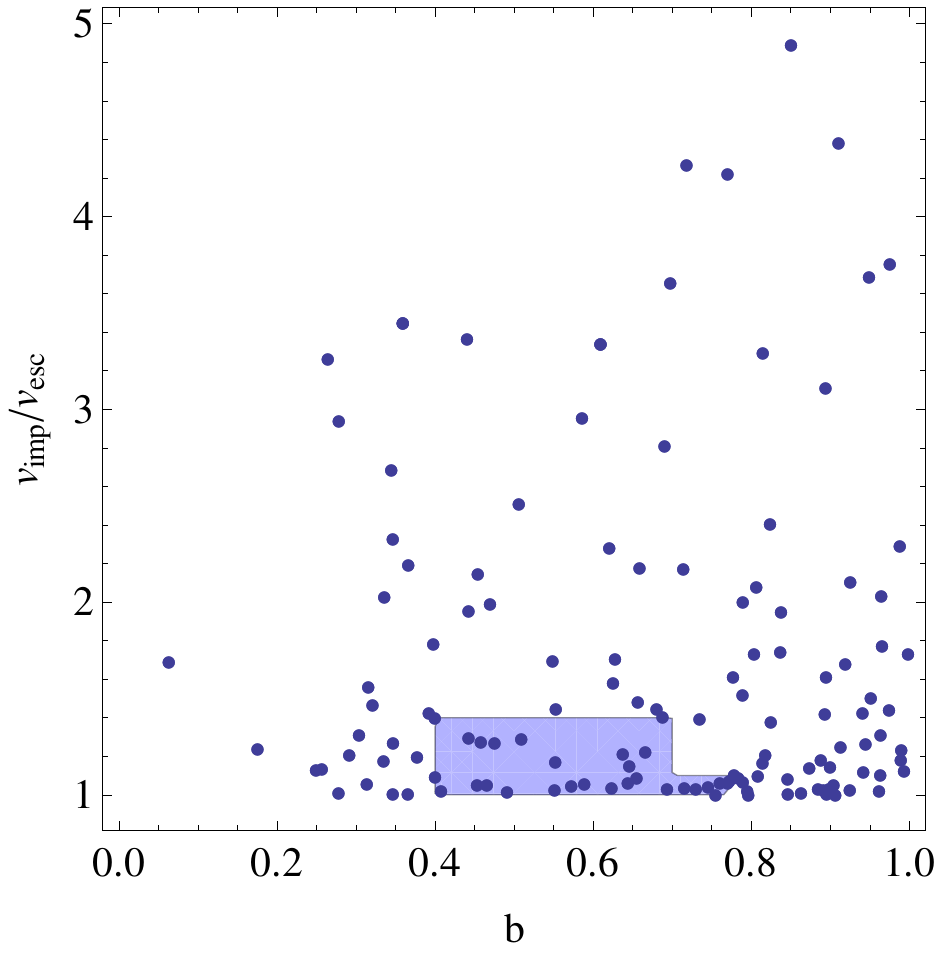}%
\caption{ The parameter space in the ($ b$, $ v_{\rm imp}/v_{\rm esc})$-plane. The darker region gives the region for with equation (\ref{eq:diskmass}) holds up to a factor of 2. The shown sample is the final sample without the constraints in the ($ b$, $ v)$-plane. Similar to figure \ref{fig:targimp}, the number of accepted events will increase when applying the cut, since an earlier moon forming impact on the planet can become suitable. As one expects, there are more collisions with larger impact parameter. Ignoring events with smaller $b$ or higher $v$ (see text), there is still a significant group of events with $b>0.8$. Much material escapes in the high velocity cases and no moon will form, but the low velocities are hard to exclude or calculate.}
\label{fig:vb}%
\end{figure}

In the (v,b)-plane, we use the same restrictive limits to constrain the collision events. Figure \ref{fig:vb} presents all moon forming events and this parameter range. Hence, this area gives just a lower bound. We see that it covers some of the most populated parts, but a significant amount of the collisions are situated outside this area. Equation (\ref{eq:diskmass}) can help to constrain the outcome of the collisions close outside of the shaded region. A collision with a small impact parameter ($ b<0.4$) will bring very little material into orbit, whatever the impactor mass involved. In this regime, we have few events with high impact velocity and even with high velocity it might be very hard to eject a significant amount of material into orbit. Hence, we exclude them from being moon forming events. For intermediate impact parameter $(0.4<b<0.8)$, there are high velocity events ($v>1.4$). Above a certain velocity threshold, depending on $b$ and $\gamma$, most material ejected by the impact will escape the system and the disk mass might be too small to form a satellite of interest. A large parameter ($ b>0.8$) describes a highly grazing collision. It is difficult to extent equation (\ref{eq:diskmass}) for larger $b$, since these collisions will probably result in a hit-and-run events for high velocities. SPH simulations \citep{Canup04,Canup08} suggest that high relative impact velocities $(v>1.4)$ will increase rapidly the amount of material that escapes the system. Nevertheless, these sets of simulations focus not on general impacts and the multi-dimensional collision parameter space is not studied well enough to describe those collisions in more detail. Detailed studies of particle collisions will hopefully be published in the near future (e.g. \citet{Kokubo10}). Based on the arguments above, the events inside of the shaded area form our final sample. Most of the collisions outside the area will not form a moon. The collisions with $0.4<b<0.8$ and velocity slightly above $v=1.4$ and with $b>0.8$ and small velocity ($v<1.4$) are events with unknown outcome. Including those events, the final number of possible moon forming events is increased by not more than a factor of 1.5.

\paragraph{Tidal evolution - Rotation period}
In order to separate receding satellites from satellites which are decelerated after their formation, we check if the initial semi-major axis of the satellite is situated outside or inside of the synchronous radius of the planet. In figure \ref{fig:rotation}, the sample is plotted twice, once including a general correction for angular momentum loss due to realistic collisions and once without correction. Moreover, two thresholds for the synchronous radius are shown. We choose the threshold at $1.3 a_{\rm R}$ (dashed line) and the corrected rotation period (full circles) to be the most justified case. There are two most extreme cases: the threshold situated at $1 a_{\rm R}$ (dotted line) and a corrected rotation period (full circles) indicates that only one in five collisions lead to receding moons. On the other hand, the threshold situated at $1.3\, a_{\rm R}$ and a rotation period directly obtained form the simulations (empty circles) indicates that only one in eight collisions lead to a non-receding moon.

\paragraph{Tidal evolution - Retrograde satellites}
To exclude retrograde orbiting satellites, we use a simple relation between the angular momenta involved in a collision. Since we have no exact data of the angular momentum distribution after the collision, this limit is very approximate. 
To get an estimate of the quality of this angular momentum argument, we study the effect of this threshold on the final sample. First, roughly half of all moon-forming impacts are retrograde. But in almost every case, $L_{\rm orbit}$ is much smaller than the initial angular momentum. Only four of the retrograde collisions have not enough angular momentum to provide a significant change of the spin axis and those  planet-satellite systems remain retrograde. Therefore, our final estimate is not very sensitive to this angular momentum argument. A larger set of particle collision simulations could provide better insight into the angular momentum distribution.

\paragraph{Collision history}
Two issues affecting the collision history can change the final result. These being the mass cut we choose to avoid resolution effects, where we exclude small impactors and targets, and the uncertainty about the final state of the planet-satellite system because of multiple and subsequent major impacts. The first limit seems to be well justified. If we change the threshold mass or exclude the cut completely, we change mainly the number of very small, non-satellite forming, impacts since in these cases 
there is usually insufficient energy to bring a significant amount of material into orbit.
The second issue has a significant effect on the final sample, however. Group $c$, the single moon forming collisions followed by large impacts, contains the most uncertain set of events: Neither all satellites survive the subsequent collisions nor is it likely that all satellites get lost. 
Hence, in the extreme case, almost half of all initially formed moons, all of Group $c$, could be lost and our final result reduced significantly.

\begin{table}[h]

\begin{tabular}[ht]{|l|c|}
  \hline
   & Uncertainty factor\\
  \hline
  \hline
  Range of initial parameters & high\\
  Number of simulations & high\\
  \hline
  Collision parameter & 1-1.5\\
  Tidal evolution - Rotation & 0.5-2\\
  Tidal evolution - Retrograde orbit & $\sim\,1$\\
  Collision history & 0.5-1\\
  \hline
  Satellite mass - disk mass & 0.5-2\\
  Satellite mass - accretion efficiency & 0.2-1\\  
  \hline

\end{tabular}

  \caption{A list of the different conditions that affect the final number of satellites. The uncertainty factor gives the range in which the final result varies around the most justified. 1 equals the final value. The first two factors can not be estimated, it shows that additional simulations would be helpful. The estimation of the satellite mass is separated from the rest of the list, since the uncertainty on this estimate does not change the number of satellites in the final sample, in contrast to the others. The exclusion of retrograde satellites is very approximate, but it has almost no effect on the final sample and therefore this factor is close to 1.}
\end{table}

An overview of the above uncertainties in given in table 1. Additional planet formation simulations are necessary to quantify a large part of existing uncertainties. Simulations of protoplanet collisions exploring the multi-dimensional collision parameter space are desirable and will hopefully be published soon and might help to constrain the parameters for moon formation better. A study of the effect of subsequent accretion of giant bodies after a moon forming impact might give better insight in the evolution of a planet-satellite system.

\section{Discussion and results} 
\label{sec:results}

\begin{figure}
\centering%
\includegraphics[scale=0.8]{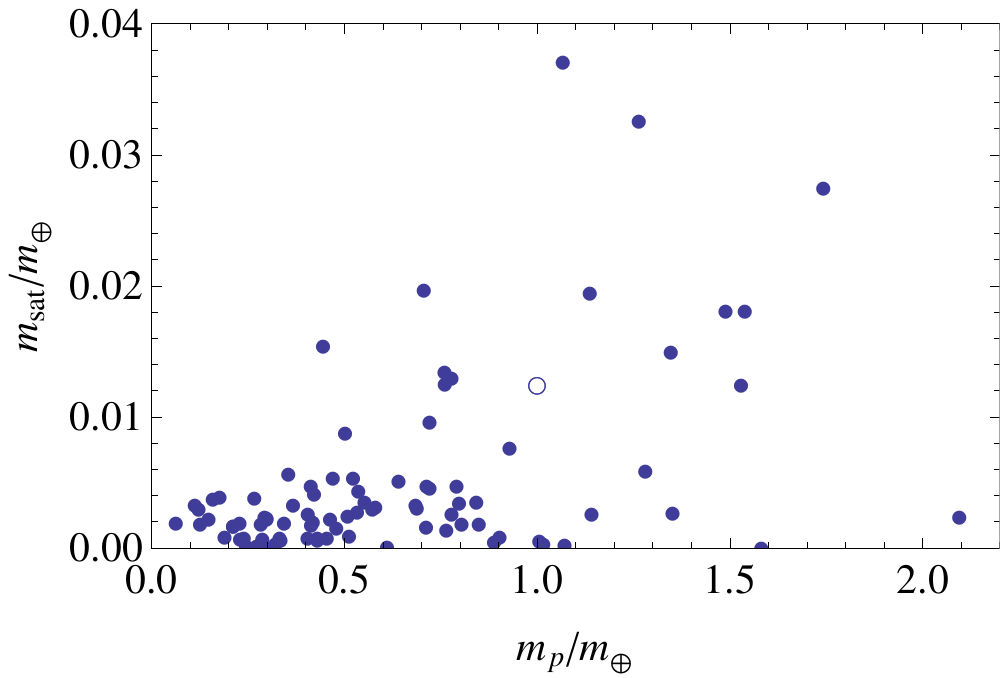}%
\caption{ The masses of the final outcomes of the planets for which we identified satellite forming collisions. $m_{\rm satellite}$ is the mass of the satellite, assuming an accretion efficiency of $50\%$, and $m_p$ is the mass of the planet after the complete accretion. The circle indicates the position of the Earth-Moon system with the assumption $m_{\rm disk}= m_{\rm Moon}$. }
\label{fig:masses}%
\end{figure}

Under these restrictive conditions we identify 88 moon forming events in 64 simulations, the masses of the resulting planet-satellite systems are shown in figure \ref{fig:masses}. On average, every simulation gives three terrestrial planets with different masses and orbital characteristics and we have roughly 180 planets in total. Hence, almost one in two planets has an obliquity stabilizing satellite in its orbit. If we focus on Earth-Moon like systems, where we have a massive planet with a final mass larger than half of an Earth mass and a satellite larger than half a Lunar mass, we identify 15 moon forming collisions. Therefore, 1 in 12 terrestrial planets is hosting a massive moon. The main source of uncertainties results from the modelling of the collision outcomes and evolution of the planet-satellite system as well as the small number of simulation and the limited range of initial conditions. We do not include the latter in our estimate. Hence, we expect the total number of Earth-Moon like systems in all our simulations to be in a range from 4 to 45. This results in a low-end estimate of 1 in 45 and a high-end estimate of 1 in 4. In addition, taking into account the uncertainties on the estimation of the satellite mass, roughly 60 of those systems are formed in the best case or almost no such massive satellites are formed if the efficiency of the satellite accretion in the circumplanetary disk is very low.

There are several papers, where the authors performed N-body simulations and searched for moon-forming collisions. \citet{Agnor99} started with 22-50 planetary embryos in a narrow disk centered at 1\,AU. They estimated around 2 potentially moon-forming collisions per simulation, where the total angular momentum of the encounter exceeds the angular momentum of the Earth-Moon system. They pointed out that this number is somewhat sensitive to the number, spacing, and masses of the initial embryos. \citet{Obrien06} performed simulations with 25 roughly Mars-mass embryos embedded in a disk of 1000 non-interacting (with each other) planetesimals in an annulus from 0.3 to 4.0\,AU. They found that giant impact events which could form the Moon occur frequently in the simulations. These collisions include a roughly Earth-size target whose last large impactor has a mass of $0.11-0.14 M_{E}$ and a velocity, when taken at infinity, of 4\,km/s, as found by \citet{Canup04}. O'Brien et al. pointed out that their initial embryo mass is close to the impactor mass. \citet{Raymond09} set up about 90 embryos with masses from $0.005$ to $0.1 M_{E}$ in a disk of more than 1000 planetesimals, again with non-interaction of the latter. Assuming again Canup's requirements ($v/v_{esc}<1.1$, $0.67<\sin\theta<0.76$, $0.11<\gamma<0.15$), only $4\%$ of their late giant impacts fulfill the angle and velocity criteria. They concluded that Earth's Moon must be a cosmic rarity but a much larger range of late giant collisions would produce satellites with different properties than the Moon. The initial embryo size seems to play a role in those results. In contrast, the simulations of \citet{Morishima10} start with 2000 fully interacting planetesimals and since 
embryos form self-consistently out of the planetesimals in these simulations, problems with how to seed embryos are completely 
avoided. To constrain the simulations, \citet{Morishima10} were interested in the timing of the Moon-forming impact. To identify potentially events, they searched for a total mass of the impactor and the target $>0.5m_{\oplus}$, a impactor mass $>0.05 m_{\oplus}$ and a impact angular momentum $>L_{\rm EM}$. They found almost 100 suitable impacts in the 64 simulations. Since their sample also includes high velocity or grazing impacts, although constrained through the more general angular momentum limit, and does not take into account collision history and tidal evolution, the difference to our result is not surprising.

Life on planets without a massive stabilizing moon would face sudden and drastic changes in climate, posing a survival challenge that has not existed for life on Earth. Our simulations show that Earth-like planets are common in the habitable zone, but planets with massive, obliquity stabilizing moons do occur only in 10\% of these.  

\section*{Acknowledgments}
We thank David O'Brien and an anonymous reviewer for many helpful comments. We thank University of Zurich for the financial support. We thank Doug Potter for supporting the computations made on zBox at University of Zurich.


\begin{thebibliography}{99}
\bibitem[Atobe et al.(2004)]{Atobe04}Atobe, K., Ida, S., Ito, T., 2004. Obliquity variations of terrestrial planets in habitable zones. Icarus. 168, 223-236 
\bibitem[Atobe and Ida(2006)]{Atobe06}Atobe, K., Ida, S., 2006. Obliquity evolution of extrasolar terrestrial planets. Icarus. 188, 1-17 
\bibitem[Agnor et al.(1999)]{Agnor99}Agnor, C. B., Canup, R. M., Levison, H. F., 1999. On the Character and Consequences of Large Impacts in the Late Stage of Terrestrial Planet Formation. Icarus. 142, 219-237
\bibitem[Berger et al.(1984)]{Berger84}Berger, A.L., Imbrie, J., Hays, J., Kukla, G., Saltzman B. (Eds.), 1984. Milankovitch and Climate - Understanding the Response to Astronomical Forcing. D. Reidel,  Norwell
\bibitem[Berger(1989)]{Berger89}Berger, A. (Ed.), 1989. Climate and Geo-Science - A Challenge for Science and Society in the 21st Century. Kluwer Academic, Dordrecht
\bibitem[Cameron and Ward(1976)]{Cameron76}Cameron, A. G. W., Ward, W. R., 1976. The origin of the Moon. Proc. Lunar Planet Sci. cd cConf. 7, 120–122
\bibitem[Cameron and Benz(1991)]{Cameron91}Cameron, A.G.W., Benz, W., 1991. The origin of the Moon and the single-impact hypothesis IV. Icarus. 92, 204-216
\bibitem[Canup and Esposito(1996)]{Canup96}Canup, M. R., Esposito, L. W., 1996. Accretion of the Moon from an Impact-Generated Disk. Icarus. 119, 427-446
\bibitem[Canup(2004)]{Canup04}Canup, M. R., 2004. Simulations of a late lunar-forming impact. Icarus. 168, 433-456
\bibitem[Canup(2008)]{Canup08}Canup, M. R., 2008. Lunar-forming collisions with pre-impact rotation. Icarus. 196, 518-538
\bibitem[Chamberlin(1905)]{Chamberlin05}Chamberlin, T. C., 1905. In Carnegie Institution Year Book 3 for 1904, 195-234, Washington, DC: Carnegie Inst.
\bibitem[Chambers and Wetherill(1998)]{Chambers98}Chambers, J. E., Wetherill, G. W., 1998. Making the Terrestrial Planets: N-Body Integrations of Planetary Embryos in Three Dimensions. Icarus 136, 304-327
\bibitem[Chambers(1998)]{Chambers99}Chambers, J. E. 1999. A hybrid symplectic integrator that permits close encounters between
massive bodies. MNRAS 304, 793-799
\bibitem[Dones and Tremaine(1993)]{Dones93}Dones, L., Tremaine, S., 1993. On the Origin of Planetary Spin. Icarus. 103, 67-92
\bibitem[Duncan et al.(1998)]{Duncan98}Duncan, M., Levison, H.F., Lee, M.H., 1998. A multiple time step symplectic algorithm for integrating close encounters. Astron. J. 116, 2067-2077
\bibitem[Goldreich and Peale(1970)]{Goldreich70}Goldreich, P., Peale, S.J., 1970. The obliquity of Venus. Astron. J. 75, 273-284
\bibitem[Halliday(2000)]{Halliday00}Halliday, A.N., 2000. Terrestrial accretion rates and the origin of the Moon, Earth and Planetary Science Letters. 176, 17-30  
\bibitem[Hartmann and Davis(1975)]{Hartmann75}Hartmann, W. K., Davis, D. R., 1975. Satellite-sized planetesimals and lunar origin. Icarus. 24, 504–515
\bibitem[Ida et al.(1997)]{Ida97}Ida, S., Canup, R.M., Stewart, G.R., 1997. Lunar accretion from an impact-generated disk. Nature. 389, 353-357
\bibitem[Kokubo and Ida(1998)]{Kokubo98}Kokubo, E., Ida, S., 1998. Oligarchic growth of protoplanets. Icarus. 131, 171-178
\bibitem[Kokubo et al.(2000)]{Kokubo00}Kokubo, E., Ida, S., Makino, J., 2000. Evolution of a Circumterrestrial Disk and Formation of a Single Moon. Icarus. 148, 419-436
\bibitem[Kokubo et al.(2006)]{Kokubo06}Kokubo, E., Kominami, J., Ida, S., 2006. Formation of terrestrial planets from protoplanets. I. Statistics of basic dynamical properties. Astrophys. J. 642, 1131-1139 
\bibitem[Kokubo and Genda(2010)]{Kokubo10}Kokubo, E., Genda, H., 2010. Formation of Terrestrial Planets from Protoplanets Under a Realistic Accretion Condition. Astrophys. J. 714, 21-25
\bibitem[Laskar and Robutel(1993)]{Laskar93a}Laskar, J., Robutel, P., 1993. The chaotic obliquity of the planets. Nature. 361, 608-612
\bibitem[Laskar et al.(1993)]{Laskar93}Laskar, J., Joutel, F., Robutel, P., 1993. Stabilization of the Earth's obliquity by the Moon. Nature. 361, 615-617 
\bibitem[Laskar(1996)]{Laskar96}Laskar, J., 1996. Large scale chaos and marginal stability in the solar system. Celest. Mech. Dynam. Astron. 6, 115-162
\bibitem[Lathe(2004)]{Lathe04}Lathe, R., 2004. Fast tidal cycling and the origin of life. Icarus. 168, 18-22 
\bibitem[Lathe(2006)]{Lathe06}Lathe, R., 2006. Early tides: Response to Varga et al. Icarus 180, 277-280
\bibitem[Lissauer(1993)]{Lissauer93}Lissauer, J. J., 1993 Planet Formation. Ann. Rev. Astron. Astrophys. 31, 129-174
\bibitem[Milankovitch(1941)]{Milan41} Milankovitch, M., 1941. Kanon der Erdbestrahlung und seine Anwendung auf das Eiszeitproblem. Kanon Koeniglich Serbische Academie Publication. 133
\bibitem[Morishima et al.(2010)]{Morishima10}Morishima, R., Stadel, J., Moore, B., 2010. From planetesimals to terrestrial planets: N-body simulations including the effects of nebular gas and giant planets. Icarus. 207, 517-535
\bibitem[Murray and Dermott(1999)]{Murray99} Murray, C.D., Dermott, S.F., 1999. Solar System Dynamics. Cambridge University Press, Cambridge	
\bibitem[O'Brien et al.(2006)]{Obrien06} O'Brien, D. P., Morbidelli, A., Levison, H. F., 2006. Terrestrial planet formation with strong dynamical friction. Icarus. 184, 39-58
\bibitem[Ohtsuki(1993)]{Ohtsuki93} Ohtsuki, K., 1993. Capture Probability of Colliding Planetesimals: Dynamical Constraints on Accretion of Planets, Satellites, and Ring Particles. Icarus. 106, 228-246
\bibitem[Raymond et al.(2004)]{Raymond04}Raymond, S. N., Quinn, Th., Lunine, J. I., 2004. Making other earths: dynamical simulations of terrestrial planet formation and water delivery. Icarus. 168, 1-17
\bibitem[Raymond et al.(2009)]{Raymond09}Raymond, S. N., O'Brien, D.P., Morbidelli, A., Kaib, N.A., 2009. Building the terrestrial planets: constrained accretion in the inner solar system. Icarus. 203, 644-662 
\bibitem[Safranov(1969)]{Safranov69}Safronov, V. S., 1969. Evolution of the Protoplanetary Cloud and Formation of the Earth and Planets. Moskau, Nauka. Engl. transl. NASA TTF-677, 1972
\bibitem[Stadel(2001)]{Stadel01}Stadel, J., 2001. Cosmological N-body simulations and their analysis. University of Washington, Ph.D.
\bibitem[Stewart and Leinhardt(2009)]{Stewart99}Stewart, S.T., Leinhardt, Z.M., 2009. Velocity-dependent catastrophic disruption criteria for planetesimals. Astrophys. J. 691, 133-137
\bibitem[Touboul et al.(2007)]{Touboul07}Touboul, M., Kleine, T., Bourdon, B., Palme, H., Wieler, R., 2007. Late formation and prolonged differentation of the Moon inferred from W isotopes in lunar metals. Nature. 450, 1206-1209
\bibitem[Varga et al.(2006)]{Varga06}Varga, P., Rybicki, K.R., Denis, C., 2006. Comment on the paper "Fast tidal cycling and the origin of life" by Richard Lathe. Icarus. 180, 274-276
\bibitem[Ward(1974)]{Ward74}Ward, W.R., 1974. Climate variations on Mars. 1. Astronomical theory of insolation. J. Geophys. Res. 79, 3375-3386
\bibitem[Ward and Rudy(1991)]{Ward91}Ward, W.R., Rudy, D.J., 1991. Resonant obliquity of Mars? Icarus. 94, 160-164
\bibitem[Ward and Brownlee(2000)]{Ward00}Ward, P. D., Brownlee, D., 2000. Rare Earth: Why Complex Live is Uncommon in the Universe. Copernicus, Springer-Verlag, New York
\bibitem[Wetherill and Stewart(1989)]{Wetherill89}Wetherill, G. W., Stewart, G. R., 1989. Accumulation of a swarm of small planetesimals. Icarus. 77, 330-357


		


















\end{thebibliography}
\end{document}